# Convolutional Neural Network Transformer (CNNT) for Fluorescence Microscopy image Denoising with Improved Generalization and Fast Adaptation

Azaan Rehman[1], Alexander Zhovmer[2], Ryo Sato[3], Yosuke Mukoyama[3], Jiji Chen[4], Alberto Rissone[5], Rosa Puertollano[5], Harshad Vishwasrao[4], Hari Shroff[6], Christian A. Combs[7], Hui Xue[1]

**Affiliations:**

(1) Office of AI Research, National Heart, Lung and Blood Institute (NHLBI), National Institutes of Health (NIH), Bethesda, MD 20892, USA
(2) Center for Biologics Evaluation and Research, U.S. Food and Drug Administration (FDA), Silver Spring, Maryland 20903, United States.
(3) Laboratory of Stem Cell and Neurovascular Research, NHLBI, NIH, Bethesda, MD 20892, USA
(4) Advanced Imaging and Microscopy Resource, NIBIB, NIH, Bethesda, MD 20892, USA
(5) Laboratory of Protein Trafficking and Organelle Biology, NHLBI, NIH, Bethesda, MD 20892, USA
(6) Janelia Research Campus, Howard Hughes Medical Institute (HHMI), Ashburn, VA, USA
(7) Light Microscopy Core, NHLBI, NIH, Bethesda, MD 20892, USA

**Contact information:**

\* Corresponding author: Hui Xue, Ph.D., National Heart, Lung and Blood Institute, National Institutes of Health, 9000 Rockville Pike, Bethesda, MD 20892, USA

(301) 827-0156      hui.xue@nih.gov

## Abstract

Deep neural networks have been applied to improve the image quality of fluorescence microscopy imaging. Previous methods are based on convolutional neural networks (CNNs) which generally require more time-consuming training of separate models for each new imaging experiment, impairing the applicability and generalization. Once the model is trained (typically with tens to hundreds of image pairs) it can then be used to enhance new images that are like the training data. In this study, we proposed a novel imaging-transformer based model, Convolutional Neural Network Transformer (CNNT), to outperform the CNN networks for image denoising. In our scheme we have trained a single CNNT based "backbone model" from pairwise high-low SNR images for one type of fluorescence microscope (instance structured illumination, iSim). Fast adaption to new applications was achieved by fine-tuning the backbone on only 5-10 sample pairs per new experiment. Results show the CNNT backbone and fine-tuning scheme significantly reduces the training time and improves the image quality, outperformed training separate models using CNN approaches such as - RCAN and Noise2Fast. Here we show three examples of the efficacy of this approach on denoising wide-field, two-photon and confocal fluorescence data. In the confocal experiment, which is a 5x5 tiled acquisition, the fine-tuned CNNT model reduces the scan time form one hour to eight minutes, with improved quality.

## Introduction

The field of microscopy is changing rapidly to include both hardware and software-based advancements that are expanding the biological imaging frontier. The ongoing advancements in fluorescence microscopy have allowed biologists to image nearly at the molecular level[1]. Similarly, developments in light-sheet microscopy have allowed for faster and more gentle imaging than ever before[2]. These transformative breakthroughs are happening alongside constant improvements in fluorescent dyes as well as the hardware and software underlying traditional imaging techniques like confocal and two-photon microscopy. Despite these remarkable progresses there remains the compromise where the temporal and spatial resolution limits the adequate collection of photons. Importantly, lower fluorescence excitation power is often necessary to avoid photobleaching or cellular phototoxicity[3]. This last point is importance for imaging live samples where fast, low light level setup is required to avoid motion artifacts while maintaining sample health.

Deep learning algorithms are currently being used for many image processing tasks including segmentation[4], super-resolution[5], and contrast generation in the label free imaging[6]. There has been a recent spate of deep learning to restore or enhance fluorescence imaging where signal-to-noise (SNR) has been degraded due to the need to optimize imaging parameters for speed or lower dosage along the lines listed above[7–10]. These algorithms use pairs of matching high and low SNR image samples to build a model to restore (denoise or enhance) low SNR images (supervised training) or use information in the low SNR images themselves (unsupervised training) for the same purpose. In most cases the best results in restoration are achieved in supervised training (Fig. 1A). For instance, using a U-net architecture, content-aware image restoration (CARE)[11] networks enhanced image resolution, denoise, and remove resolution anisotropy. The three-dimensional residual channel attention networks (3D-RCAN)[12] have provided further gains in image enhancement for volumetric time-lapse imaging (confocal and super-resolution) and in expansion microscopy.

Although CNN models like CARE and 3D-RCAN are very effective, they require considerable time and data to train. The main limitation is the lack of fast adaptation, where new models are trained from the scratch for new experiments with lengthy computing time and a dedicated training dataset. The model trained under a combination of cell types, hardware and imaging protocol is not robustly transferrable to other experiments[12]. This limits the adaptability to new experiments for which they have not been trained. In contrast, blind zero-shot denoisers including Noise2Void and Noise2Noise and more recently Noise2Fast, have been shown to denoise images with a per-sample training[13–16], where a full training is conducted for the very low SNR image to be processed, without requiring a pre-acquired training dataset. Limitations of these methods include the slow inference due to per-sample training and inferior performance compared to supervised training[15]. The high-quality data can be degraded with realistic and diverse noise distribution[17] to mitigate the requirement of acquiring corresponding low quality data.

The transformer architecture is capable of learning long-range signal coherence and is scalable to large-scale datasets[18]. It is the foundation for the very successful large language model, such as the ChatGPT[19]. For the domain of imaging, the transformer based models show improved adaptation than CNNs as the attention mechanism, the key component in transformer, computes the data specific coefficients to transform inputs into outputs, while the CNNs applied the fixed parameter sets after training to all input data[20]. This adaptation ability is further enhanced by the multi-head soft attention heads in the transformer[18].

To improve the model adaptation for microscopy denoising experiments, we propose a novel transformer architecture which we term convolutional neural net transformer (CNNT) to effectively process large microscopy images Instead of training individual models for each experiment, we train a generalized backbone model using diverse datasets followed by fine tuning this generalized backbone for each new experiment (Fig. 1B). We show the CNNT fine-tuning for each new experiment requires only few numbers of image pairs (e.g. 5-10) which greatly reduces the time necessary to train (<10mins) and offers better performance than CNN models as well as training from the scratch. The per-trained model is generalized across different imaging modalities (e.g. from iSim to two-photon), cell types and imaging protocols. We first trained the backbone model on U2OS cells image data acquired with only an iSim scanner. The trained backbone was fine-tuned on data from different sample types imaged on iSim, wide-field, confocal and two-photon scanners. We compared the CNNT performance with RCAN3D and Noise2Fast for both image quality and training speed. Further, the live zebrafish was images in an 5x5 tile setup for large field-of-view. Low SNR images from fast imaging due to the need to freeze sample motion was significantly improved after model inference, leading to a seven-fold speedup.

## Materials and Methods

### CNNT: a novel imaging transformer model

CNNs excel at processing imagery data due to their nature of spatially invariant inductive bias. However, CNNs struggle to capture long range signal correlation along the time or Z dimension without significant computation overhead[21]. The standard transformer model performs very well on time series data due to the attention mechanism that is able to combine all time points to compute the output. However, it struggles with images due to the heavy computational costs especially when working with 2D+T or 3D images. The standard transformer works with linear attention mechanism that maps the input to key, query, value (k,q,v) matrices and then to compute the attention score. The inputs are mapped to the k,q,v matrices with linear layers, which allows for expressive learning of representations but also incurs a heavy quadratic cost with respect to the input length[22]. We introduce a novel transformer architecture that can work efficiently with 2D+T and 3D images. In our approach we opt to use convolution layers instead of linear layers to map the inputs. Not only does this reduce the complexity from quadratic to linear number of operations, but also reintroduces the spatial invariant of convolutions to the transformer architecture. This novel architecture, coined Convolution Neural Net Transformer (CNNT), allows for several enhancements from both CNNs and transformers alone.

As shown in Fig. 2, with the ability to work with flexible dimension sizes as input and output, the CNNT becomes a plug and play module which we use to create complete architectures. As a baseline we create a standalone module of CNNT cell which consists of an input projection that expands or contracts the channel dimension, followed by a CNNT attention module that enriches the input with convolved attention, followed by a standard CNN mixer that helps in sharing information across feature channels. The modular structure of CNNT cell becomes the building block for a complete model. We chose U-net *as* our base architecture as it is effective to combine information from different resolution levels and computational more efficient. This CNNT U-net has the first two levels downscaling the image spatially and increasing the feature dimension. Last two levels scale up the image spatial dimensions back to the original size and reduce the feature dimension correspondingly. Before each upscale level, we concatenate the input with the output from the corresponding down scale level. Each level of the U-net is made of four CNNT cells stacked on top of each other. We refer to the complete architecture as CNNT U-net.

**Backbone training and finetuning with CNNT U-net**

The Training was split into two stages for backbone learning and fine-tuning. Instead of training a model for every experiment, we first train a general backbone with data from several organelles pooled together. For fast adaptation to a new experiment, a finetuning step is performed, requiring only a small dataset and only a few minutes of extra train time. We found backbone trained with one microscopy type (iSim[23] in this study) is fine-tuned very well on other scope types (e.g. widefield, confocal, two-photon). The backbone training does take longer to complete, but once trained, it can be shared for many new experiments. The finetuning takes only a few minutes as the model only needs to adapt to the small new data, not learn from scratch.

We defined a backbone dataset created with a combination of 7 different organelles and 4 finetuning datasets created of 4 different organelles. The 7 different organelles used in backbone training were all under an iSim scanner with a total of 154 z-stack images for 7 different organelles. Finetuning experiments were from three different microscopies. None of the finetuning data was present in the backbone training. In each case we see a significant boost in SNR compared to the noisy raw image. In addition, CNNT also outperforms the current state-of-the-art. For each downstream task, 3 different models were finetuned using either 5, 10 or 20 samples.

**CNN models for comparison**

We compare our results with RCAN3D which is considered state-of-the-art supervised machine learning model, and Noise2Fast which is one of the best self-supervised machine learning model. RCAN3D was trained from scratch following the publication [ref]. Noise2Fast learns noise prediction for each example it sees.

RCAN[12]

RCAN 3D was trained with the code available online (https://github.com/AiviaCommunity/3D-RCAN). Following the recommendation in its paper, all data available for each downstream task was used to train RCAN3D models. The evaluation was done on the same test images as the CNNT. The training and evaluation wall-clock time duration was recorded.

Noise2Fast[15]

Noise2Fast implementation was from its official repository (https://github.com/jason-lequyer/Noise2Fast.git). Noise2Fast does not require paired samples; instead, a model is trained from scratch for every noisy image it sees. The published training parameters were used in all experiments

**Downstream finetuning tasks**

**Wide-Field Imaging of MEF Cells**

Non-muscle myosin 2A-GFP mouse embryonic fibroblast (MEF) cells were maintained in DMEM media (Gibco, Cat# 11965-092) containing 5% FBS (Gibco, Cat# 16000-044). Primary mouse non-muscle myosin 2A-GFP T cells were isolated with EasySep™ Mouse T Cell Isolation Kit, following manufacturer's instructions (Stemcell, Cat# 19851). Mouse T cells were maintained in RPMI 1640 media (Gibco, Cat# 11875093) supplemented with 10% fetal bovine serum (Gibco, Cat#16000-044) and interleukin-2 (Stemcell, Cat# 78081.1).

For this experiment, cells were cultured in glass-bottom dishes (MatTek, Cat# P35G-1.5-20-C). For immunostaining, we used β-actin (A5441, Sigma), NM2A (909801, BioLegend), Alexa Fluor 568 goat anti-mouse (ThermoFischer Scientific, Cat# A-11004), and Alexa Fluor 488 goat anti-rabbit (ThermoFischer Scientific, Cat# A78953) antibodies. For immunostaining, MEF cells were fixed and permeabilized with 4% paraformaldehyde (Sigma-Aldrich, Cat# 158127) and 0.05% Triton X-100 (Sigma-Aldrich, Cat# T8787) in PBS (KDMedical, Cat# RGF-3210) for 15 min at room temperature. To block nonspecific binding, samples were washed with 1x Blocker BSA in PBS twice, blocked for 1 hour at room temperature, and subsequently stained with 1:500 dilution of primary antibodies in 1x Blocker BSA (Thermo Scientific, Cat# 37525) in PBS for overnight at 4°C. Samples were washed three times (5 min) with 1x Blocker BSA in PBS, and stained with 1:500 secondary antibodies in 1x Blocker BSA in PBS for 2 hours at RT. Samples were washed three times (5 min) with 1x Blocker BSA in PBS. The stained samples were mounted and imaged using 90% Glycerol (Sigma-Aldrich, Cat# G2025) in PBS.

Live cell fluorescence imaging was performed using a Leica DMi8 microscope equipped with 100x/1.4 NA oil immersion objective lens and Okolab stage top incubator with CO2, temperature, and humidity control. For widefield fluorescence microscopy, low and high signal-to-noise ratio image pairs of T cells were acquired as follows: i) 25 ms exposure, 10% illumination intensity, 5% laser power; ii) 25 ms exposure, 30% illumination intensity, 5% laser power; iii) 100 ms exposure, 100% illumination intensity, 10% laser power.

**Two-photon imaging of zebrafish embryos**

All zebrafish experiments were performed in compliance with the National Institutes of Health guidelines for animal handling and research using an Animal Care and Use Committee (ACUC) approved protocol H-0252(R5). Zebrafish were raised and maintained at the temperature of 28.5C. Zebrafish handling, breeding and staging were performed as previously described (Kimmel et al., 1995. Developmental Dynamics 203:253-310; The Zebrafish Book: A Guide for the Laboratory Use of Zebrafish Danio (" Brachydanio Rerio"), M Westerfield - 2007 - University of Oregon). To prevent pigmentation, the embryos used for confocal analysis were cultured in fish water containing 0.003% 1-phenyl-2-thiouera (PTU, Sigma-Aldrich, P7629) from 24 hpf. The following strain was used: Tg(ins:dsRed)$^{m1081}$;Tg(fabp10:dsRed;ela3l:GFP)$^{gz12}$; Tg(ptf1a:EGFP)$^{jh1}$ transgenic line (Anderson et al., "Loss of Dnmt1 catalytic activity reveals multiple roles for DNA methylation during pancreas development and regeneration", Developmental Biology, Volume 334, Issue 1, 1 October 2009, Pages 213-223). Originally, the line was obtained crossing the Tg(ins:dsRed)$^{m1081}$;Tg(fabp10:dsRed;ela3l:GFP)$^{gz12}$ line, also known as 2-Color Liver Insulin acinar Pancreas (2CLIP) with the Tg(ptf1a:EGFP)$^{jh1}$ line.

Imaging of zebrafish embryos

Two-photon imaging of zebrafish embryos was performed on Tg(ins:dsRed)$^{m1081}$;Tg(fabp10:dsRed;ela3l:GFP)$^{gz12}$; Tg(ptf1a:EGFP)$^{jh1}$ transgenic line at room temperature using a LEICA SP8 confocal microscope and a 25x (0.95 NA) water dipping lens (Leica HC FLUOTAR L VISIR) with a dual beam Insight (Ti:sapphire) laser (Newport/Spectra-Physics, Irvine, CA) . The fish express dsRED fluorescent protein in the islets of Langerhans in the endocrine pancreas, driven by the insulin (*ins*) promoter and in the liver hepatocytes, driven by the fatty acid binding protein 10a (*fabp10a* gene). Heterozygous parents were crossed and then the collected embryos were selected at 3 days post fertilization (dpf) using a fluorescent SteREO Discovery.V12 stereomicroscope (Zeiss). At 5 dpf, zebrafish embryos were anesthetized using a buffered tricaine methanesulfonate (MS-222, Sigma-Aldrich, E10521) solution in 0.003% PTU solution in E3 medium. The anesthetized embryos were then included in a 1% solution of low melting agarose dissolved in 0.003% PTU solution in E3 medium on a glass coverslip (Warner Instruments, CS-40R15) and carefully oriented in a lateral position. During the whole imaging embryos were kept in a solution of MS-222 and 0.003% PTU in E3 medium.  Pairs of low-noise "ground-truth" and fast "noisy" image stacks  of DsRed and GFP were acquired at a scan rate of 8000 Hz using a resonant scanner with a format of 512 x 512 pixels, 0.2 x 0.2 micron pixel sizes, and excitation at 1045nm (DsRed) and 920nm (GFP), with emission bandwidths of 650-700nm(DsRed) and 500-552nm (GFP), and an interslice distance of 0.5 microns. A line average of 8 was used for the ground truth images whereas no line averages were used for the noisy image stacks resulting in a decreased time of imaging of more than 6 folds.

The scanners allow us to perform repeated imaging and average them to reduce the noise. We exploit this to create a series of images with consistently increasing SNR by taking 64 repetitions of the same field-of-view and then averaging the first $n$ time points to get the $n^{th}$ image in the series. We selected time average 1-63 as the noisy image and average 64 as the ground truth. We test the finetuned model on all time point averages from 1 to 64 to evaluate model behaviors for inputs with different SNR levels.

**Confocal Imaging of Mouse Lung Tissue**

Lungs isolated from C57BL/6 mice were inflated with 4% paraformaldehyde/PBS and immersed in 4% paraformaldehyde/PBS at 4°C overnight. After fixation, the lungs were immersed in 30% sucrose/PBS at 4°C overnight and then embedded in OCT compound. Cryosectioning of the lung tissues was performed at a thickness of 50 μm and mounted on Superfrost Plus Gold microscope slides (Fisher). Immunostaining was performed with the following primary antibodies:

rat anti-CD45 antibody (1:500, eBioscience, 14-0451-85), mouse anti-αSMA-Cy3 (1:500, Millipore Sigma, C6198), armenian hamster anti-PECAM-1/CD31 antibody (1:300, Chemicon, MAB1398z), and syrian hamster anti-Podoplanin-APC (1:50, Biolegend, 127410). For immunofluorescent detection, donkey anti-rat IgG-Alexa 488 (1:250, Invitrogen, A21208) and goat anti-armenian hamster IgG-Cy3 (1:250, Jackson ImmunoResearch, 127-165-160) secondary antibodies were used. Nuclei were visualized with Hoechst 33342 (1:500, Biotium, 40046). Mouse lung tissue was imaged using a Leica SP8 confocal microscope in the resonant scanning mode (8000Hz) using a 63x/1.4 NA oil objective (Leica HC PL APO CS2) and the LASx software (version 3.5.7). 5x5 tiled, five-color, z-stack images were acquired with 144nm pixels in the XY dimension with a format of 1024 x 1024, an interslice distance of 2$\mu$m, with a pinhole set to 1 A.U. Excitation and emission ranges for Hoechst, Alexa Fluor 488, CY3, PECAM, and APC were 405 with 415-458, 500 with 509-542, 550 with 559-587, 594 with 606-635, and 650 with 660-745nm respectively. Line-average varied with "ground truth" images collected with a line average of 32 while faster "noisy" test images were collected with a line average set to be 4. The ground-truth images were acquired in ~1hour, but noisy data was acquired in ~8mins.

**Results**

Instead of training new models for every experiment from scratch, a general backbone was trained and then finetuned with a few new samples to quickly adapt to a new experiment. In each downstream task we finetuned the backbone with 5, 10 or 20 samples and compared the CNNT results with RCAN3D and Noise2Fast. In all tasks the CNNT considerably improves the SNR and outputs consistently good results no matter the number of samples used to finetune. Models show robust adaptation across scanner types, samples imaged and acquisition condition.

The finetuning time for CNNT is less than 10 minutes for 10 samples, as seen in Fig 3-6. The inference time is a minute on big images of size of 100x1200x1200. RCAN has a slower train time of 2 hours using the shared codebase and configuration of large model. Noise2Fast has to learn noises for every data, leading to every long inference time.

**Widefield**

The MEF cells are imaged with a widefield camera. For this modality 20 images were collected and 10 were separated for testing. 5 and 10 samples were individually used to finetune the CNNT. RCAN was trained on 10 samples as well.

As shown in Fig. 3, CNNT performs well under this scenario as well and removes noise uniformly throughout the field-of-view. RCAN performs well in the center regions but does not remove noise uniformly throughout the image. Noise2Fast struggles to output sharp and clean images. The CNNT results have the higher PSNR and SSIM3D metrics, over the RCAN and Noise2Fast. The fine-tuning time is ~13 minutes for CNNT with 10 samples, while the RCAN training time is over 2 hours. Supplement Video1 shows the MEF cell images before and after CNNT model finetuned with 10 samples, and the ground-truth data for the reference.

**Two-photon**

For this experiment we curated 24 train images focusing on zebrafish liver and pancreas. 6 separate images of full scans were collected as test set. Out of 24 train images, CNNT was trained on either 5, 10 or 20 samples. RCAN was trained on all 24.

Fig. 4 shows CNNT outputs clean images with rich details, even surpass the quality of the ground truth. This may indicate extra quality gain can be achieved with pre-training. For different number of fine-tuned samples, CNNT image quality remain steady, better than RCAN3D and Noise2Fast. Due to the ground-truth images are noisy, we notice CNNT does not give the highest PSNR. The fine-tuning time is ~6mins for 10 samples, much faster than training from the scratch. Supplement Video 2 shows zebra fish pancreas results, against the ground-truth.

The time point average experiment was devised to find the breaking point of CNNT U-net. By computing averages for different number of time points, input images with different level of signal-noise-ratio are created. These set of averaged images are inputted into CNNT to test its robustness.

Fig. 5 gives model outputs for averages 1, 4, 8, 16, 32, and 64. We see that CNNT U-net is robust against this wide range of input SNRs. For Avg 1 where the input signal is weak, the model recovers the detailed structures. With improved input SNR for more averages, the model behaved robustly, giving consistently good quality.

**Confocal**

The final experiment is to image the mouse lung tissue with a confocal scanner for 5 different colors. In this experiment we push the limits and see how much time saving can be achieved with CNNT. As seen in Fig. 6, the input image misses structural information due to fast imaging, while the ground truth shows well connected boundaries. There is a total of 245 images consisting of 49 images from 5 different channels. 25 images, or 5 images per channel were separated for testing. From remaining 220 images CNNT was trained on 5, 10 or 20 randomly selected samples. RCAN was trained on all 220 images. Supplement Video 3 renders first three channels as RGB colors to compare 8mins scans before and after CNNT model and one-hour long ground-truth scan.

As shown in Fig. 6, the CNNT recovers the image details from rather sparsely sampled data. It indicates the backbone training allows the model to learn general structure information of samples as well as the denoising. CNNT offers high PSNR and SSIM3D than RCAN and Noise2Fast, even the RCAN was trained with a lot more samples. Noise2Fast struggles to produce very clean outputs. The fine-tuning is much faster in training time, ~7mins for 10 samples, vs. over 2 hours training from the scratch.

**Discussion**

In this paper we propose a novel Convolutional Neural Network Transformer and a new training method for the light microscopy denoising. With these contributions, we pushed the boundaries of microscopy image scanning in speed and quality. The backbone model was trained on iSim datasets and fine-tuned on wide-field, two-photon and confocal microscopies. Results show the CNNT architecture and pre-training lead to improved image quality, with much less training time, compared to training from the scratch. It outperforms the convolution-based models and allow fast adaptation to new imaging experiments.

We also studied the robustness of CNNT, via the multiple averaging tests. Results show the finetuned CNNT model gave consistently good quality outputs, even for the Avg 1 input with lower SNR. In practice, it is plausible to acquire a dataset with multiple repetitions and determine the maximal speedup in imaging by checking the outputs with varied input SNRs to find out the model break point. Given the faster finetuning time of CNNT method, a new experiment can first determine the optimal imaging setup by quickly training new models from the backbone and then commence the bulk imaging job.

One limitation of training from the scratch is the model cannot utilize data acquired from other imaging sessions. The backbone training is capable of consuming large amount of curated data. Based on the scaling law[24], the performance can be further improved as the amount of trained data and model sizes. This can lead to the light microscopy specific foundation model for imaging, if a sizable backbone model can be pre-trained on a very large and diverse dataset. Resulting models may benefit both imaging and downstream analysis of light microscopy images.

**Conclusion**

In this paper we introduced a novel deep learning architecture: Convolutional Neural Net Transformer, a hybrid architecture of CNNs and transformers and used it to denoise low SNR microscopy images. CNNT improves over standard CNNs due to the ability to work on arbitrary time or depth dimension using the attention mechanism. We further improved model adaptation by utilizing a backbone and finetuning scheme. With backbone model trained on iSim datasets, the CNNT is successfully finetuned on widefield, two-photon and confocal microscopies, with improved image quality and much faster training time.

## Conflict of Interest

The authors declare no conflict of interest.

## Data Availability

Training and test datasets for widefield, two-photon and confocal experiments are published on the NIH data repo (link). The RCAN3D iSim datasets was shared with its original publication.

## Code Availability

The CNNT model and training code used in this study are available at https://github.com/AzR919/CNNT.git. An installation guide, data and instructions for use are also available from the same webpage.

## Acknowledgements

This research was supported by the intramural research programs of the National Institute of Heart, Lung, and Blood, and the National Institute of Biomedical Imaging and Bioengineering within the National Institutes of Health.

## Author contributions

H. Xue, A. Rehman and C. Combs conceived the project and designed the experiments. H. Xue and A. Rehman developed the CNNT model and programed the training software. A. Rehman performed the model training, comparison experiments and data analysis. J. Chen, H. Vishwasrao, H. Shroff supported the RCAN3D testing and acquired the iSim datasets. R. Sato, Y. Mukoyama acquired the mouse lung tissue datasets. A. Rissone, R. Puertollano and C. Combs acquired the zebra fish data. A. Zhovmer acquired the MEF cell datasets. All authors contributed to draft the manuscript and approve the final submission.

**Figure Legends**

**Fig. 1 | Backbone and finetuning to train the light microscopy image enhancement model.** A) A separate model is trained for every imaging experiment. This training from the scratch method proves to be effective but need more samples and longer training time to reach good performance. Furthermore, since every training is independent, the model under training cannot utilize the curated datasets to help current imaging experiment. B) We propose first to train a backbone model from the large, diverse and previously curated datasets. The trained backbone model is further finetuned for every new experiment, with much smaller amount of new data. Given an effective backbone model architecture, this method will be much faster in training, and also allow the current training to reuse information acquired in previous experiments. Inspired by the success of transformer model in language pre-training, we proposed a novel imaging transformer architecture, CNNT, to serve as an effective backbone.

**Fig. 2 | The CNNT-Unet architecture.** A) The whole model consists of a pre and post convolution layers and the backbone. The input tensor has the size of [B, C, Z, H, W] for batch, channel, depth, height and weight. The C input channel are first uplifted to 32 input channels into the backbone. The post-conv layer will convert the output tensor from the backbone to C channel. There is a long-term skip connection over the backbone. B) The backbone has a Unet structure, consisting of two downsample blocks and to upsample blocks. Every downsample CNNT block will double the number of channels but reduce the spatial size by x2. Every upsample block will reduce the number of channels and expand the spatial size. C) The CNNT block includes only the CNNT cell. Every cell contains CNN attention, instance norm and CNN mixer. This design mimics the standard transformer cell design but replacing the linear attention and mixer to be the CNN attention and CNN mixer, to reduce computational cost for high resolution images. D) The CNN attention is the key part of imaging transformer cell. Unlike the linear layers in the standard transformer, the key, value and query tensors are computed with convolution layers, which

reduces the computation to process the high-resolution images and also keep the good inductive bias. The attention coefficients are computed between query and key and applied to value tensor to compute attention outputs.

**Fig. 3 | The Widefield experiment to image the MEF cells.** The pre-trained CNNT backbone was finetuned on 5 and 10 widefield image samples. The resulting model was compared to RCAN3D and Noise2Fast for image quality and computing time. A) The low-quality noisy image as the input to the models. B-C) The CNNT results after finetuning for 30 epochs on 5 and 10 samples. The quality improvement is noticeable. E) The RCAN3D model trained from the scratch for 300 epochs gave good improvement. F) The Noise2Fast result is subpar. G) the high-quality ground-truth for SSIM3D and PSNR computation and for reference. The CNNT finetuning is much faster than RCAN3D training and Noise2Fast and offer better quality measurements.

**Fig. 4 | The two-photon experiments for the pancreas of a zebra fish.** A) The low-quality image does to provide enough SNR and contrast to delineate features like islets. B, C, D) The CNNT greatly improved the image quality for 5, 10 and 20 samples. The model is robust for even 5 samples, leading to a very fast ~3.5mins finetuning time. E and F) The RCAN3D and Noise2Fast training are much longer with suboptimal quality recovery. G) The ground-truth in this experiment bears a still lower SNR. The models achieved better quality than the ground-truth images.

**Fig. 5 | The multi-average tests for the zebra fish imaging.** The imaging was repeated for N=64 times. CNNT models were tested for robustness for different level of input quality with increasing number of averages. The Avg 1, 4, 8, 16, 32 and 64 were shown here. The model was found to be robust against the lower input SNR, giving consistently boost of image quality. The model also robustly recovered finer features. No sign of hallucination was found.

**Fig. 6 | The confocal imaging for the mouse lung tissue.** A) The low-quality image was acquired with very low photon counts. B, C, D) CNNT finetuning with 5, 10, 20 samples shows recovered tissue structures and removal of background random noise. E) The RCAN3D model also gave good improvement in quality. F) The Noise2Fast had more signal fluctuation, compared to supervised models. G) The high-quality ground-truth reveals the tissue anatomical structure. Again, the timesaving of CNNT finetuning is prominent, with superior or similar image quality.

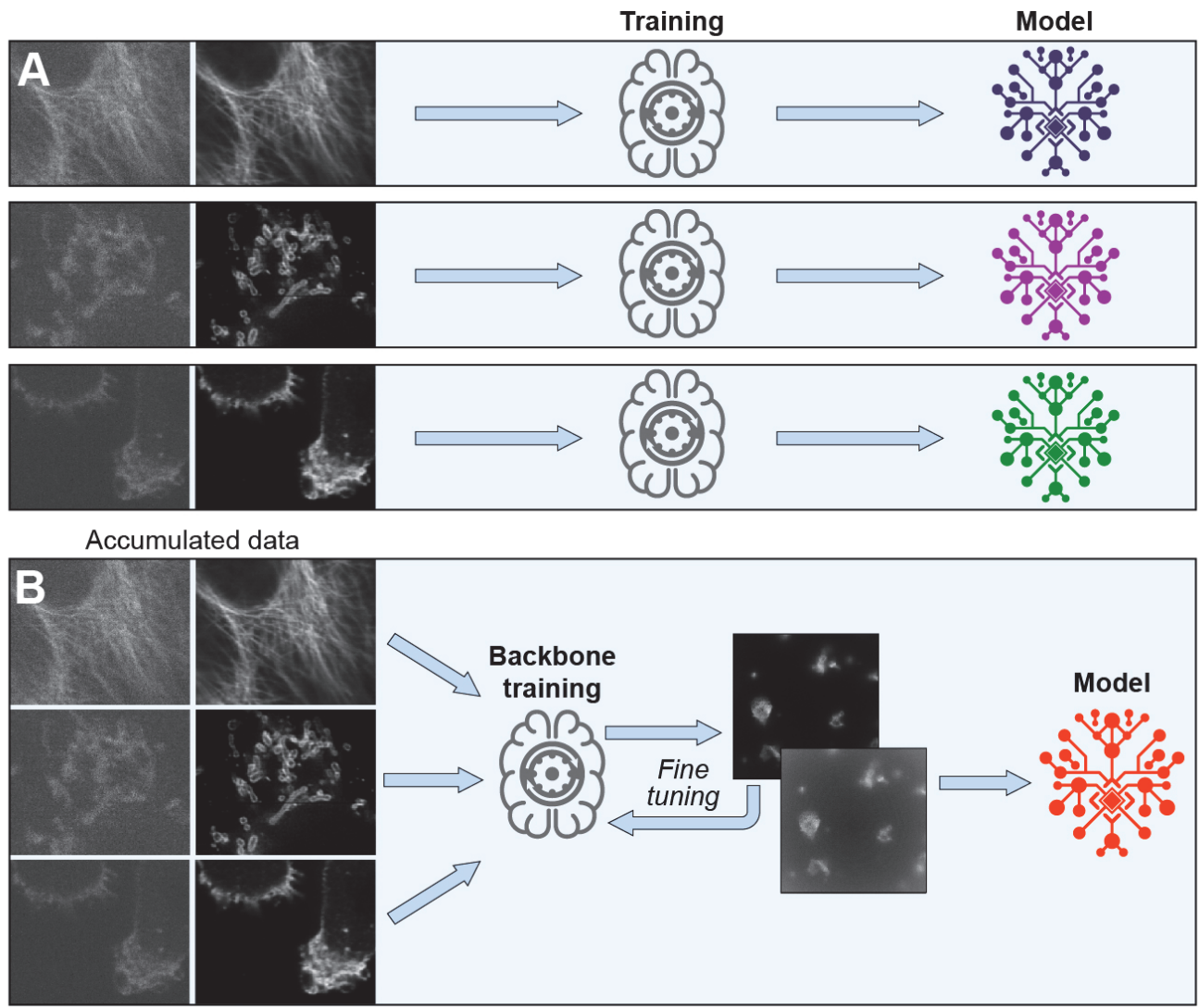

Fig. 1 | Backbone and finetuning to train the light microscopy image enhancement model. A) A separate model is trained for every imaging experiment. This training from the scratch method proves to be effective but need more samples and longer training time to reach good performance. Furthermore, since every training is independent, the model under training cannot utilize the curated datasets to help current imaging experiment. B) We propose first to train a backbone model from the large, diverse and previously curated datasets. The trained backbone model is further finetuned for every new experiment, with much smaller amount of new data. Given an effective backbone model architecture, this method will be much faster in training, and also allow the current training to reuse information acquired in previous experiments. Inspired by the success of transformer model in language pre-training, we proposed a novel imaging transformer architecture, CNNT, to serve as an effective backbone.

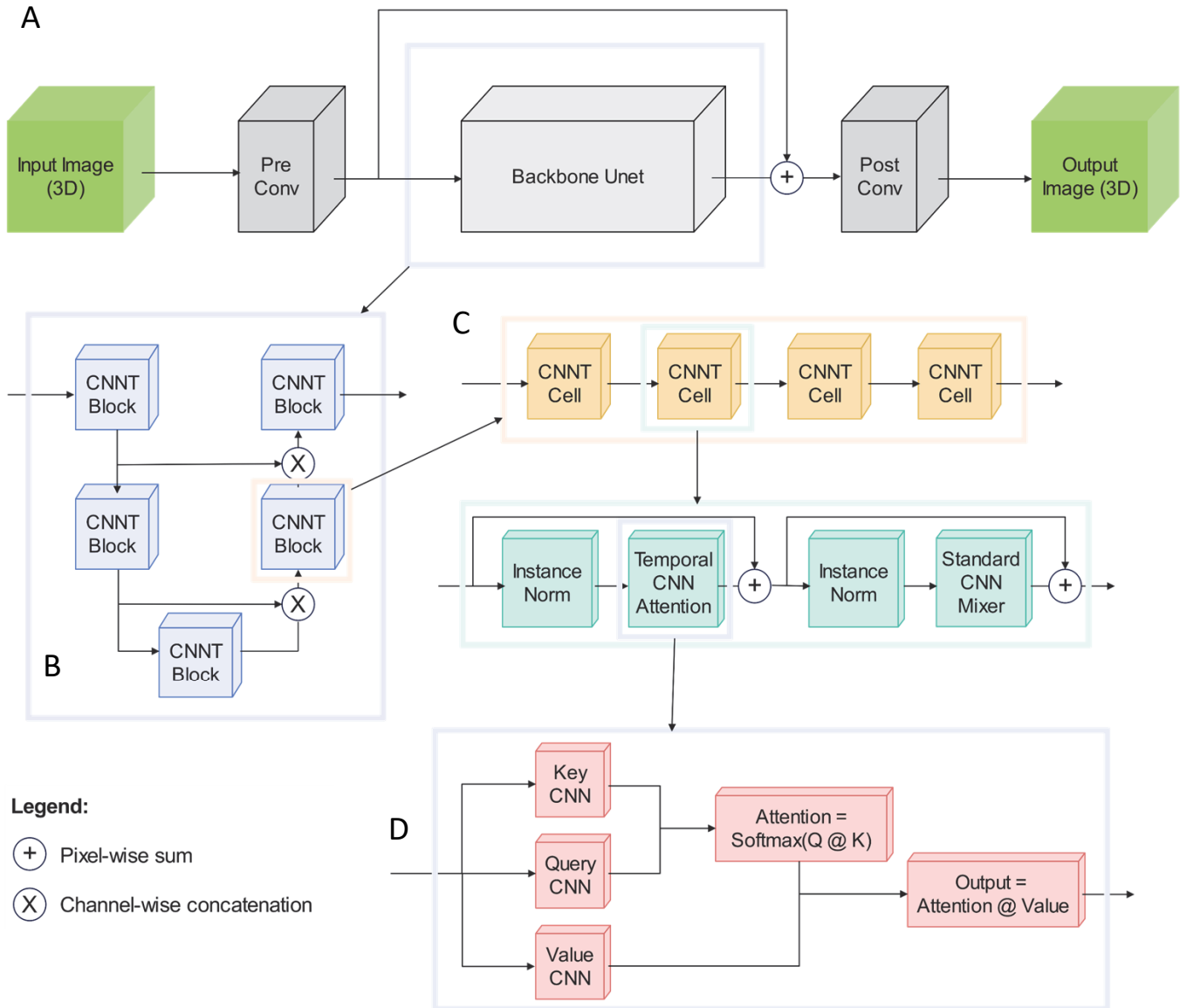

Fig. 2 | The CNNT-Unet architecture. A) The whole model consists of a pre and post convolution layers and the backbone. The input tensor has the size of [B, C, Z, H, W] for batch, channel, depth, height and weight. The C input channel are first uplifted to 32 input channels into the backbone. The post-conv layer will convert the output tensor from the backbone to C channel. There is a long-term skip connection over the backbone. B) The backbone has a Unet structure, consisting of two downsample blocks and to upsample blocks. Every downsample CNNT block will double the number of channels but reduce the spatial size by x2. Every upsample block will reduce the number of channels and expand the spatial size. C) The CNNT block includes only the CNNT cell. Every cell contains CNN attention, instance norm and CNN mixer. This design mimics the standard transformer cell design but replacing the linear attention and mixer to be the CNN attention and CNN mixer, to reduce computational cost for high resolution images. D) The CNN attention is the key part of imaging transformer cell. Unlike the linear layers in the standard transformer, the key, value and query tensors are computed with convolution layers, which reduces the computation to process the high-resolution images and also keep the good inductive bias. The attention coefficients are computed between query and key and applied to value tensor to compute attention outputs.

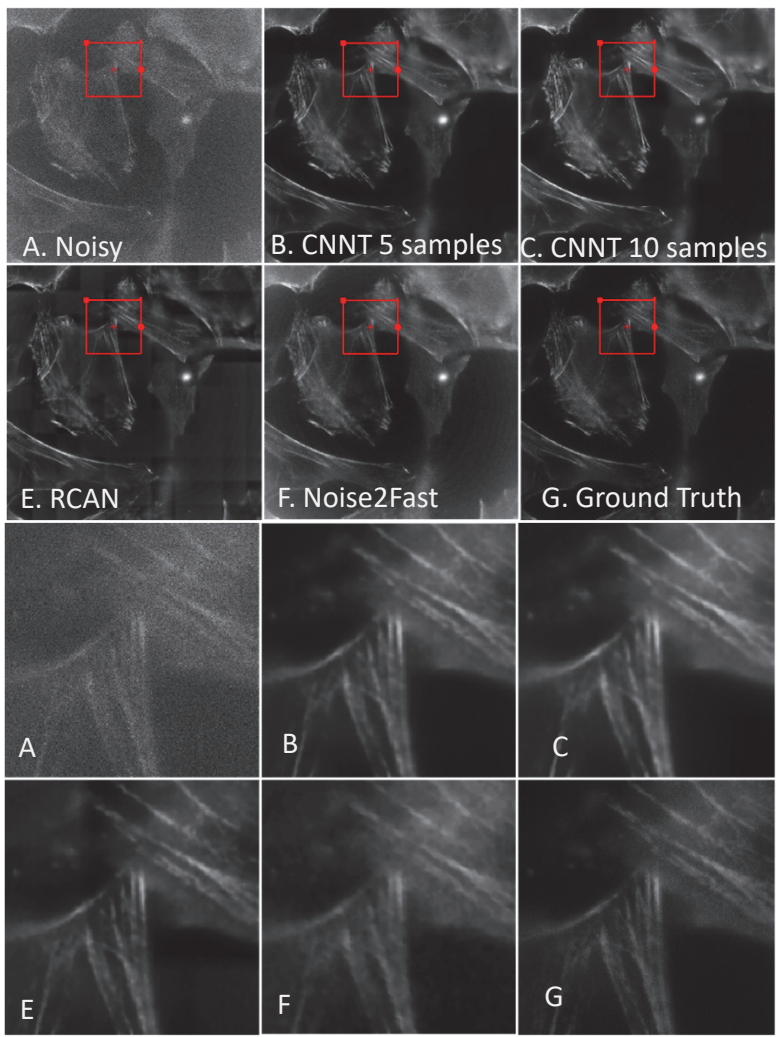
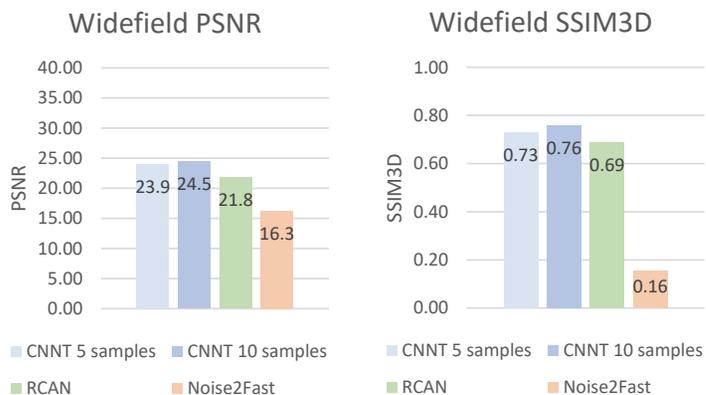

| Number of test inference images | 10 |
| --- | --- |
| Average image size | 101x1200x1200 |

| Model | Train time | Avg eval time per im |
| --- | --- | --- |
| CNNT 5 samples | 293s (4m 53s) | 68s (1m 08s) |
| CNNT 10 samples | 776s (12m 56s) | 69s (1m 09s) |
| RCAN3D | 7856s (2h 10m 56s) | 48s (48s) |
| Noise2Fast | 0s (0s) | 3220s (53m 40s) |

**Fig. 3 | The Widefield experiment to image the MEF cells.** The pre-trained CNNT backbone was finetuned on 5 and 10 widefield image samples. The resulting model was compared to RCAN3D and Noise2Fast for image quality and computing time. A) The low quality noisy image as the input to the models. B-C) The CNNT results after finetuning for 30 epochs on 5 and 10 samples. The quality improvement is noticeable. E) The RCAN3D model trained from the scratch for 300 epochs gave good improvement. F) The Noise2Fast result is subpar. G) the high quality ground-truth for SSIM3D and PSNR computation and for reference. The CNNT finetuning is much faster than RCAN3D training and Noise2Fast and offer better quality measurements.

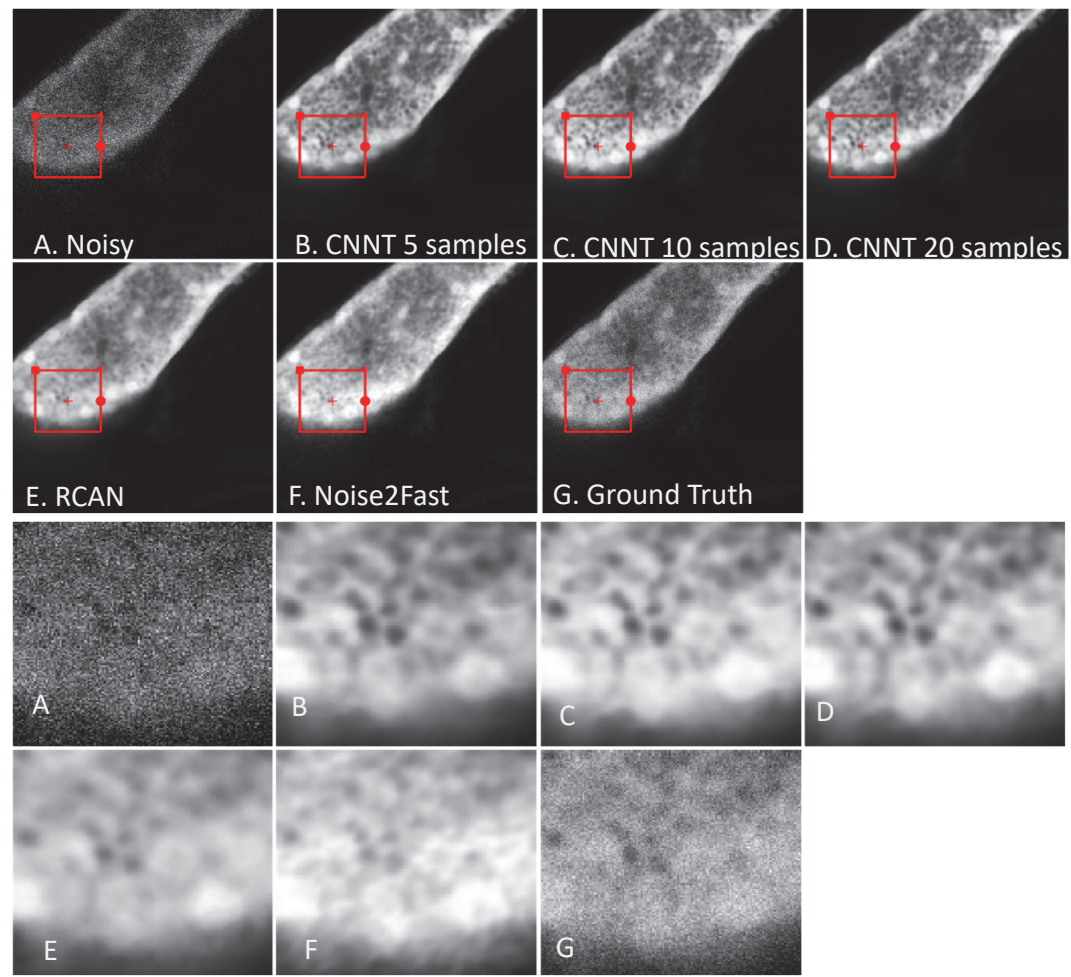
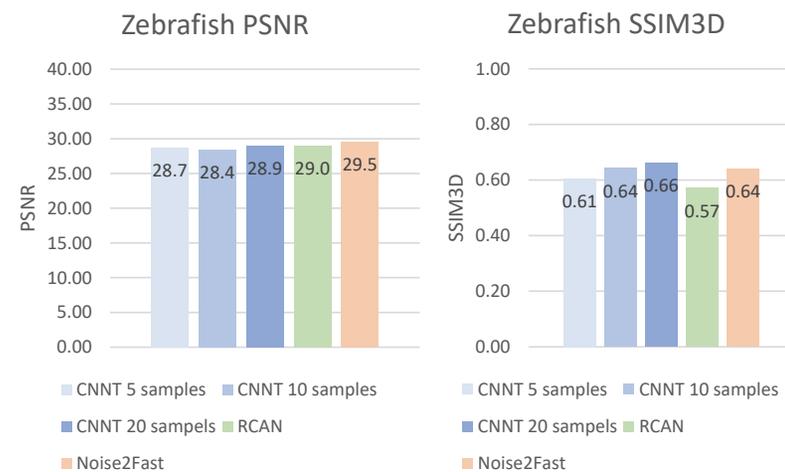

Fig. 4 | The two-photon experiments for the pancreas of a zebra fish. A) The low quality image does to provide enough SNR and contrast to delineate features like islets. B, C, D) The CNNT greatly improved the image quality for 5, 10 and 20 samples. The model is robust for even 5 samples, leading to a very fast ~3.5mins finetuning time. E and F) The RCAN3D and Noise2Fast training are much longer with suboptimal quality recovery. G) The ground-truth in this experiment bears a still lower SNR. The models achieved better quality than the ground-truth images.

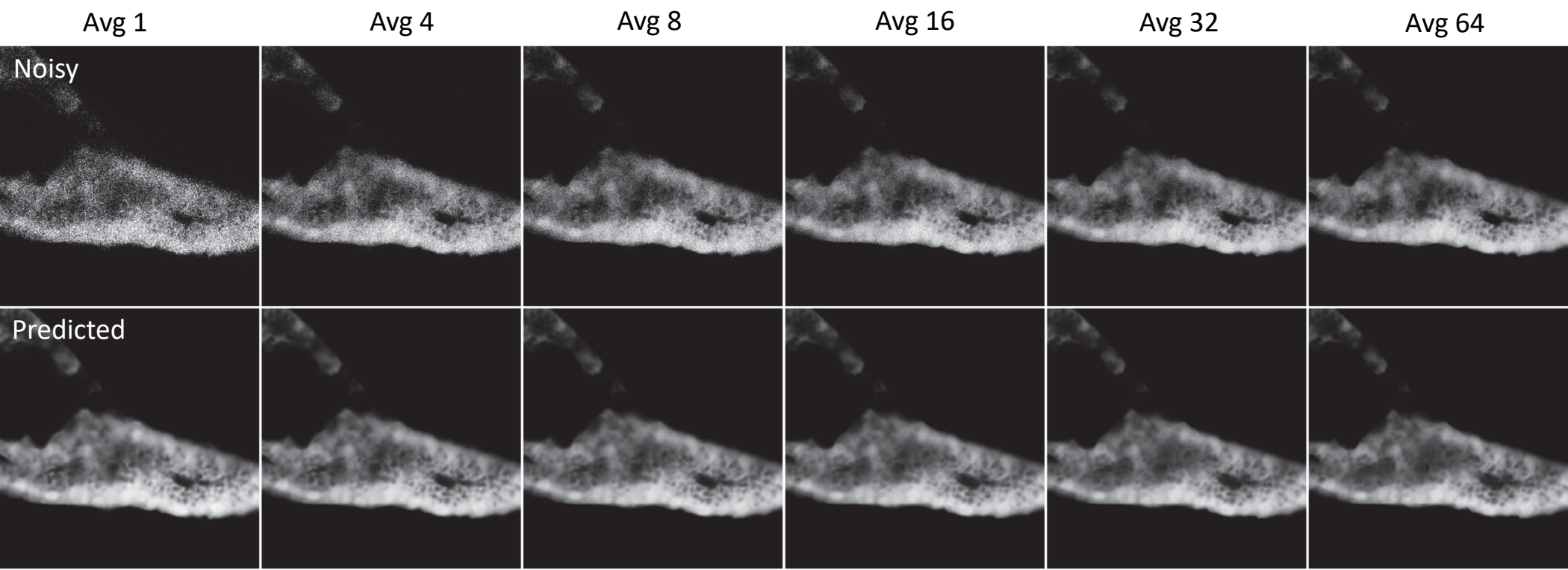

Fig. 5 | **The multi-average tests for the zebra fish imaging.** The imaging was repeated for N=64 times. CNNT models were tested for robustness for different level of input quality with increasing number of averages. The Avg 1, 4, 8, 16, 32 and 64 were shown here. The model was found to be robust against the lower input SNR, giving consistently boost of image quality. The model also robustly recovered finer features. No sign of hallucination was found.

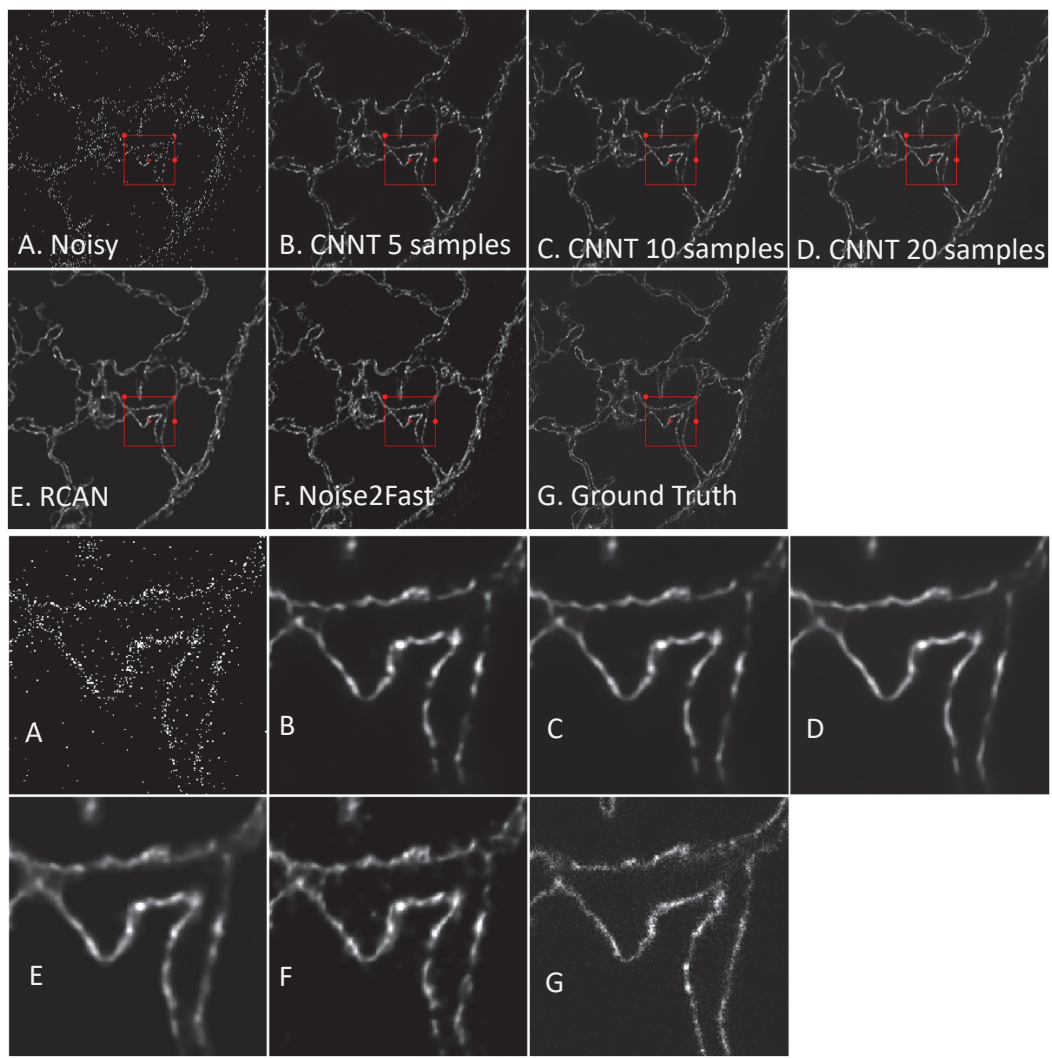
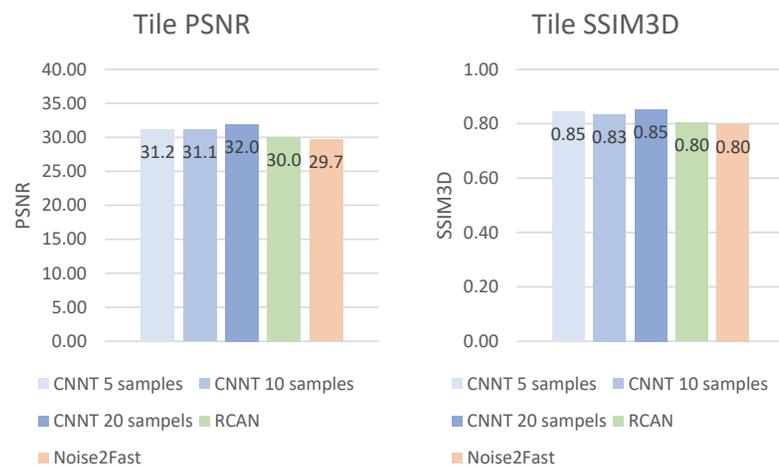

**Fig. 6 | The confocal imaging for the mouse lung tissue.** A) The low-quality image was acquired with very low photon counts. B, C, D) CNNT finetuning with 5, 10, 20 samples shows recovered tissue structures and removal of background random noise. E) The RCAN3D model also gave good improvement in quality. F) The Noise2Fast had more signal fluctuation, compared to supervised models. G) The high-quality ground-truth reveals the tissue anatomical structure. Again, the time-saving of CNNT finetuning is prominent, with superior or similar image quality.

Supplement Video 1

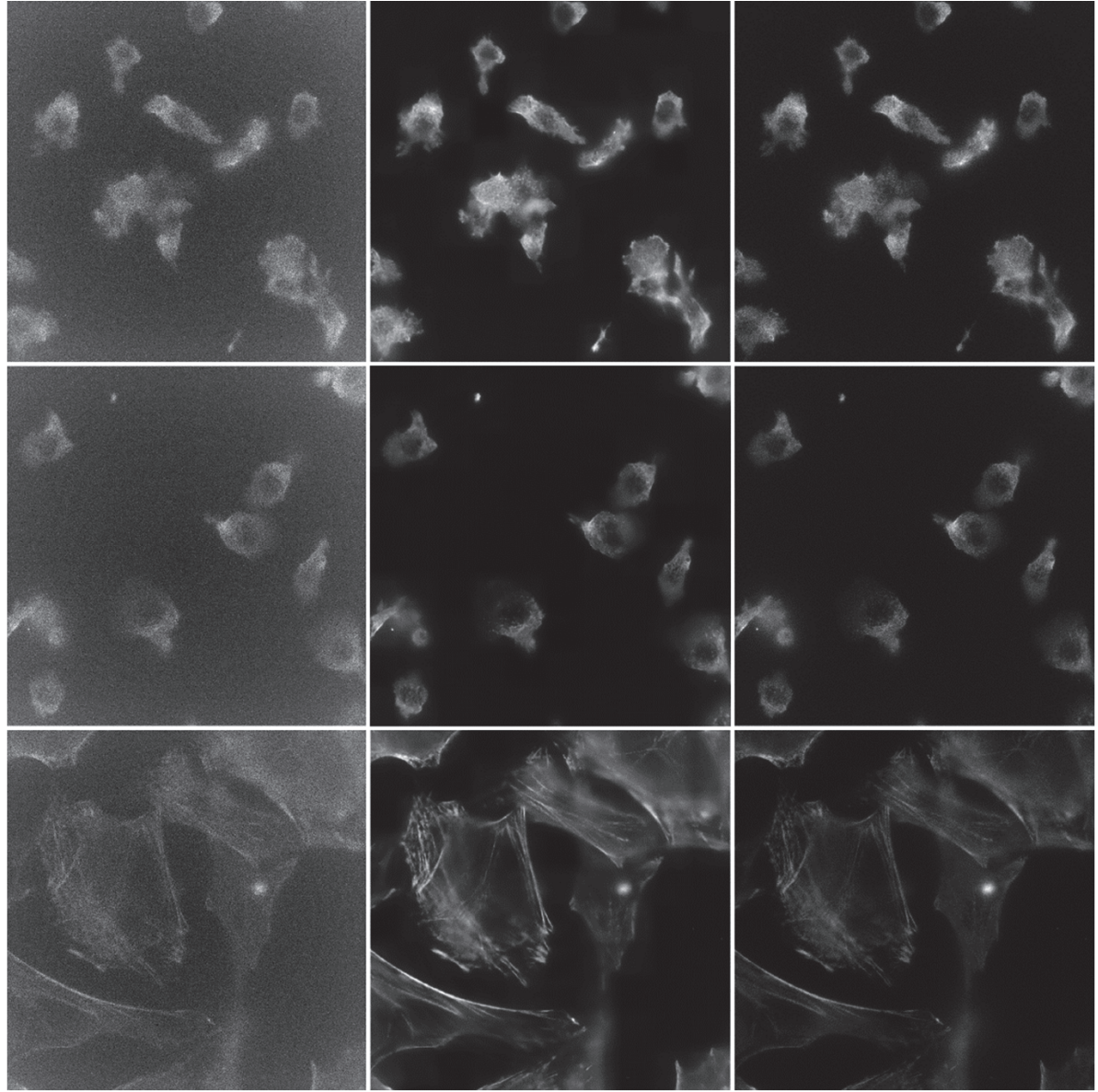

Supplement Video 2

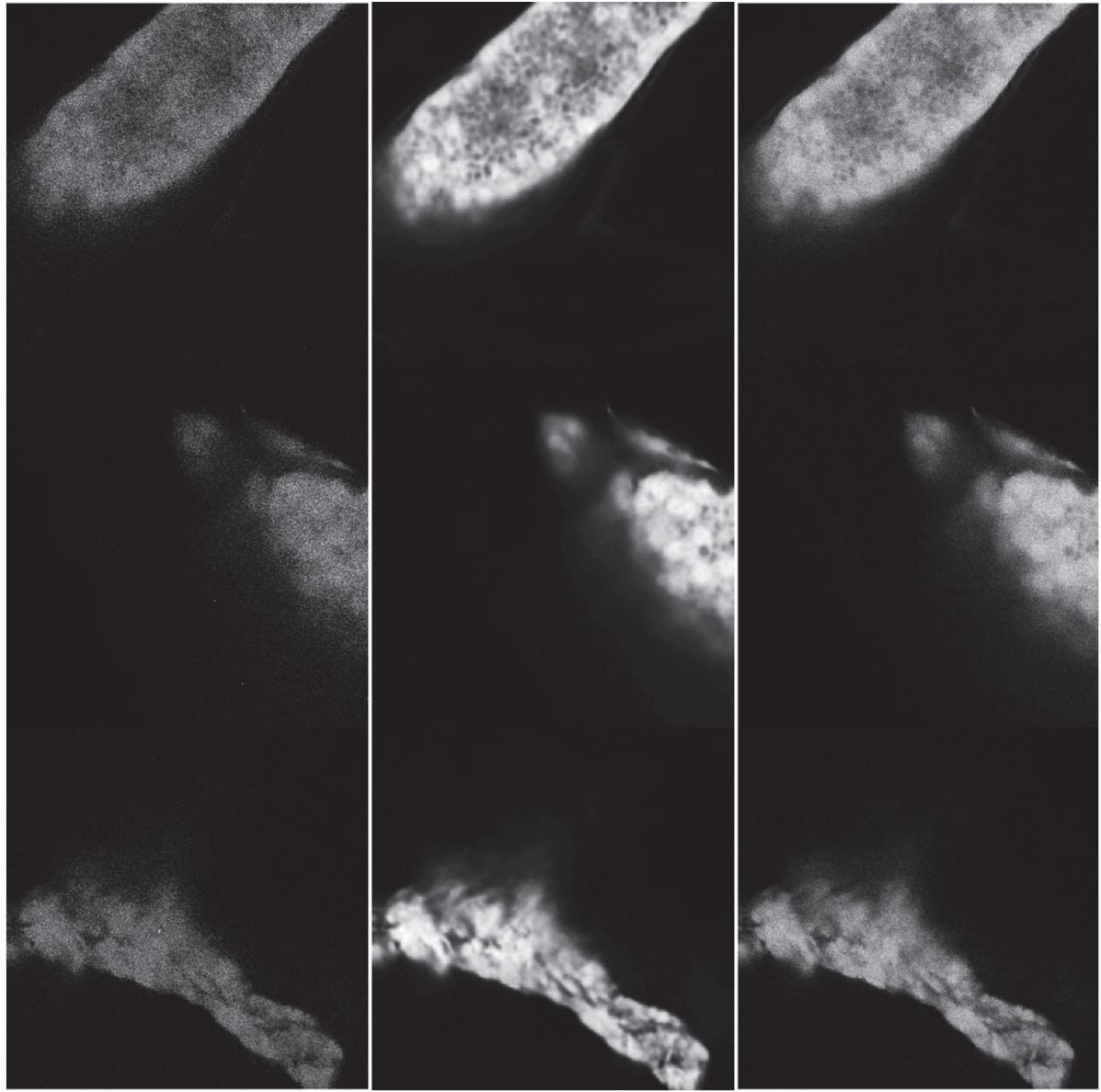

Supplement Video 3

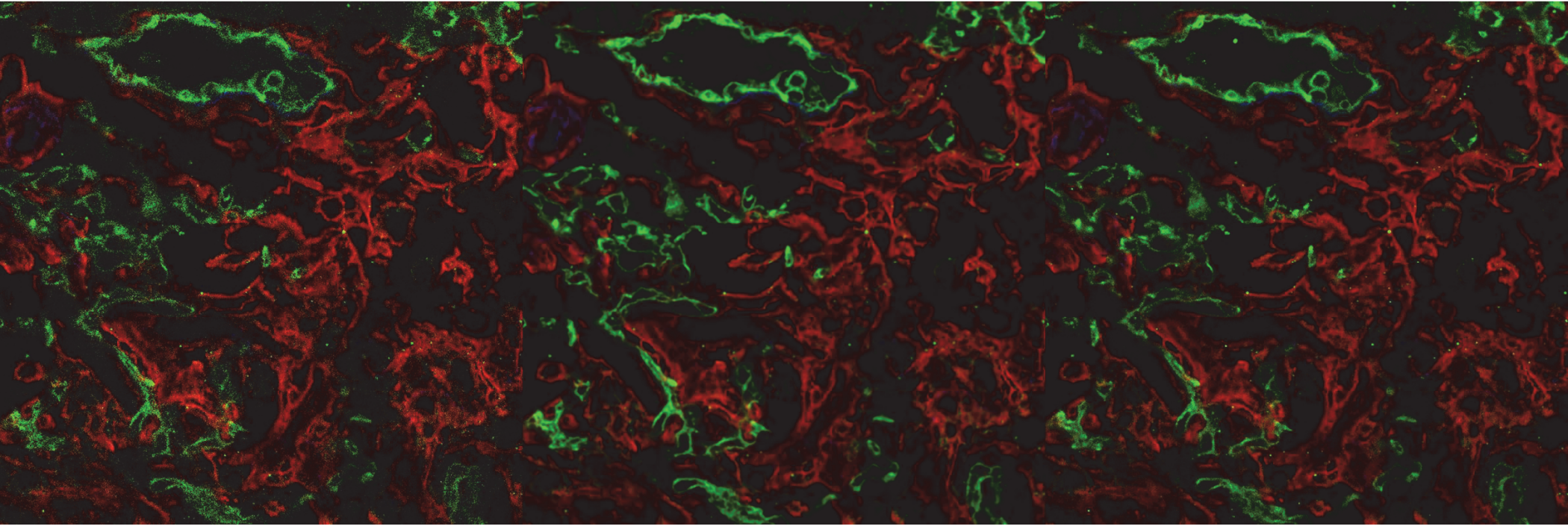